\documentclass[tradiabstract,rnote]{aa}

\usepackage{graphicx}
\usepackage{txfonts}
\usepackage{url}
\usepackage{multirow}

\begin{document}

\title{The multicomponent model of the Crab Pulsar \\ at energies above 25 GeV}

\author{	Riccardo Campana
          	\inst{1,2},
		Enrico Massaro
		\inst{1},
 		Teresa Mineo
		\inst{3},
		Giancarlo Cusumano
		\inst{3}
            }
             
\offprints{~~~~~~\\
Riccardo Campana: \texttt{riccardo.campana@uniroma1.it}}

\institute{	Department of Physics, University of Rome ``La Sapienza'', 
Piazzale A. Moro 2, I-00185, Roma, Italy
\and INAF--IASF-Roma, Via Fosso del Cavaliere 100, I-00133, Roma, Italy 
\and INAF--IASF-Palermo, Via U. La Malfa 153, I-90198, Palermo, Italy
               }

\date{Received ..., accepted ...}
\titlerunning{The GeV emission from the Crab Pulsar in the MCCM model}
\authorrunning{R. Campana et al.}
	
%
\abstract{
We review a proposed multicomponent model to explain the features of the 
pulsed emission and spectrum of the Crab Pulsar, on the light of the recent 
detection of pulsed emission above 25 GeV from the MAGIC atmospheric \v{C}erenkov 
telescope.
This model explains the evolution of the pulse shape and of the phase-resolved 
spectra, ranging from the optical/UV to the GeV energy band, on the assumption 
that the observed emission is due to several components, which have spectra 
modelled as log-parabolic laws.
We show that the new MAGIC data are well consistent with the prevision of our model.
}



\maketitle

\section{Introduction}
Phase resolved energy distributions of pulsars' signals 
are important to obtain information on the 
radiation processes and geometry of the emission regions in the 
magnetosphere.
At $\gamma$-ray energies, at variance with other energy bands, the 
three brightest sources (Vela, Crab and Geminga) show remarkably 
similar pulse structures with two main peaks at a large phase 
separation ranging from 0.4 to 0.5.
This pattern is confirmed by the very recent discovery of pulsed 
emission from PSR~J2021+3651 (Halpern et~al. 2008), that has a peak 
separation of 0.47.
The double peak structure of the Crab Pulsar (PSR B0531+21) is well 
observed across the entire electromagnetic spectrum, but the intensity 
ratio between the two peaks changes with energy.
In particular, in the X and soft $\gamma$-ray ranges the emission of the 
second peak (P2) becomes higher than the first one (P1), and a relevant 
emission from the region between the two peaks (interpeak or bridge, Ip) 
increases like P2. 
This behaviour is observed up to a few MeV, where the pulse shape turns 
almost sharply to be similar, although not equal, to the optical light 
curve. 

On the basis of a large collection of data, covering 
the frequency interval from the optical to the GeV band, 
we proposed a model (Massaro et al. 2006a, hereafter MCCM) able to
describe the spectral and phase distributions by means of a double 
two-component model.
The energy spectra of these components are not described by a simple 
power law, because of the continuous spectral steepening towards high 
energies.
We found that a very satisfactory model is a parabolic law in a 
double-logarithmic plot, corresponding to a log-normal spectral 
distribution.

A useful test to verify the goodness of this multicomponent model, and 
in particular the existence of the two high energy components, is the 
study of the pulse shape at energies higher than a few GeV, where P2 
is expected to be again the dominant feature as in hard X/soft $\gamma$ 
rays.
Some hints in this direction are given by the EGRET pulse profile 
(Thompson 2004), but statistics above 5 GeV are so poor that no firm 
conclusion can be obtained.
Very recently, the MAGIC telescope (e.g. Lorenz 2004) has detected 
pulsed emission from the Crab (Aliu et al. 2008), 
at energies above 25 GeV and with 6.4 standard deviations significance,
showing the very well established two peak profile, with similar 
amplitudes of the two peaks.

In this Note we compare the MAGIC data to the predictions of the MCCM
model, and use them to constrain the high energy components.
The model is synthetically described in Sect. 2, the extension to MAGIC 
results is presented in Sect. 3 and some hypotheses about the origin of 
these components on the basis of some recent works on the high energy 
emission from young pulsars are discussed in Sect. 4.

\section{The multicomponent model of Crab pulsar }

\begin{figure*}
\includegraphics[width=0.5\textwidth]{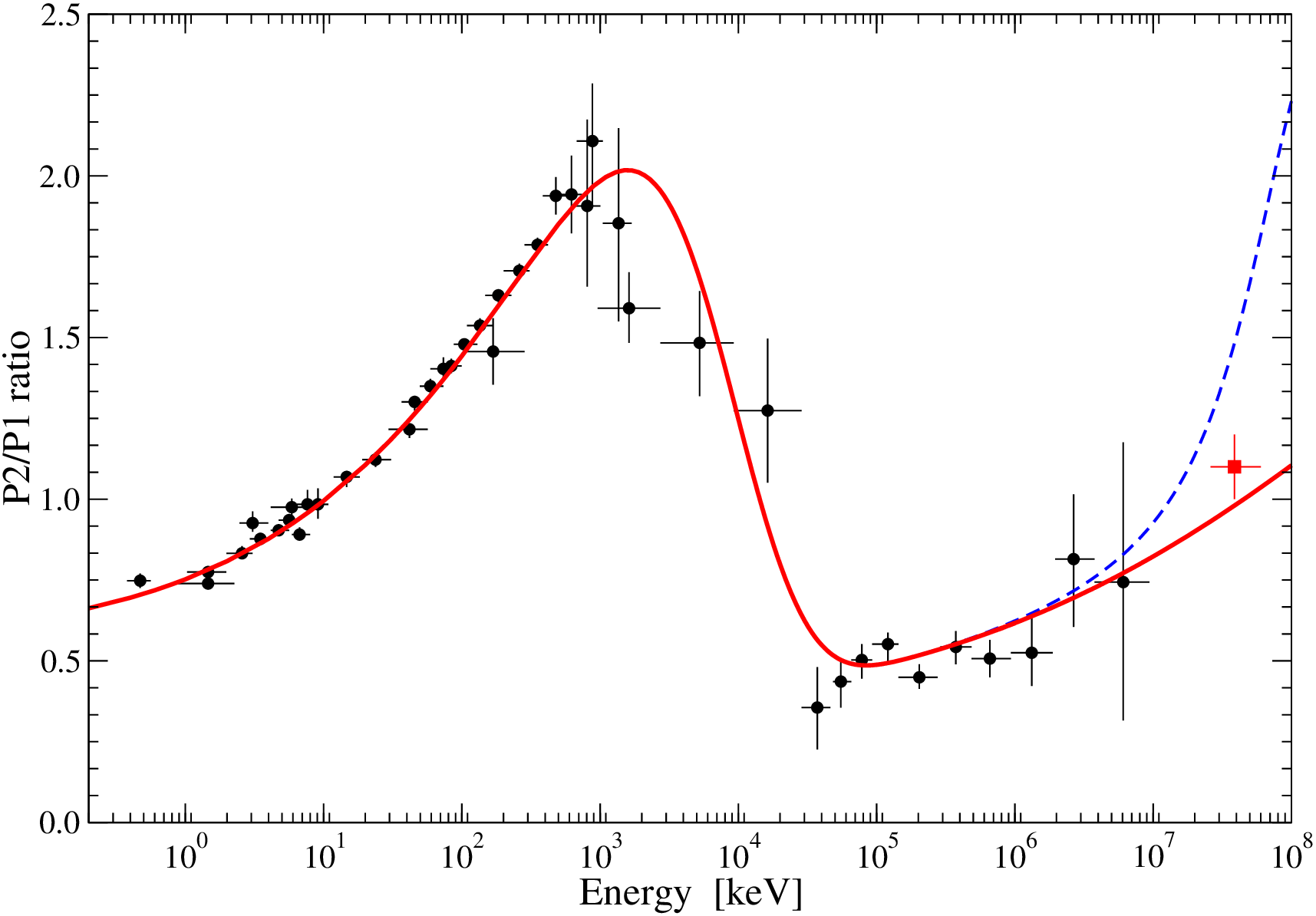}
\includegraphics[width=0.5\textwidth]{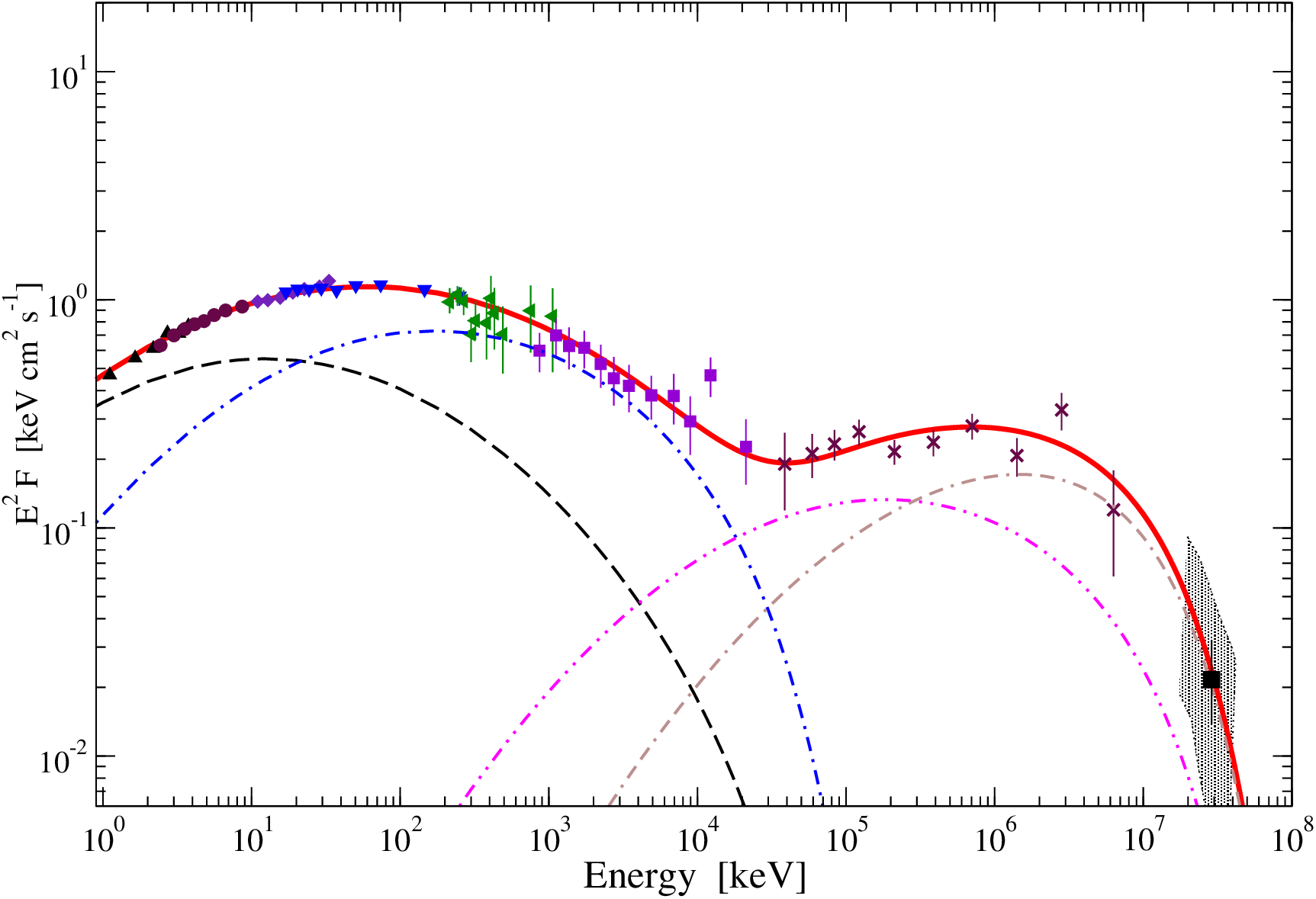}
\caption{
\emph{Left panel:} The ratio between the fluxes of P2 and P1 phase regions (P1: 
-0.06--0.04; P2: 0.32--0.43), compared to the predictions of the model. 
The black data points come from various experiments (Kuiper et al. 2001), 
while the red point at GeV energies has been obtained from MAGIC data.
The two extrapolations above 1 GeV correspond to different values 
of the cut-off energy of the $C_{O\gamma}$ spectrum: 15 GeV (solid red 
line) and 11 GeV (dashed blue line). The $C_{X\gamma}$ cut-off energy 
is fixed at 15 GeV.
\emph{Right panel:} Broadband spectral energy distribution of the total averaged pulse 
and of the four components:
dashed line: $C_{O}$; dash-dotted line: $C_{X}$; dot-dot-dashed line: 
$C_{O\gamma}$; dash-dash-dotted line: $C_{X\gamma}$.
Data points and SED of individual components are the same of the
MCCM model with the addition of the new MAGIC data 
(black square, where the shaded area represents the systematic error on the spectral fits performed by Aliu et al., 2008).}
\label{figure1}
\end{figure*}

\subsection{The two-component model: optical to hard X-rays}
 
As early presented in Massaro et~al. (2000), the spectral and phase
changes of Crab X-ray pulse shape are well reproduced by two components. 
The first component, called $C_O$, is assumed to have the same pulsed 
profile observed at optical frequencies, while the second component, 
$C_X$, is described by an analytical model to reproduce the observed 
pulse profiles. 
The latter component dominates at the interpeak (Ip) and second peak 
(P2) phase regions.
This choice can be justified also from theorical models like that proposed
by Eastlund et~al. (1997), who considered the synchrotron emission from 
marginally clamped electrons in a shell at the boundary of the closed 
magnetosphere and calculated pulse profile very similar to $C_O$. 

By means of spectral fits of BeppoSAX data, we obtained that the $C_O$ and $C_X$ components 
were described by a log-parabolic spectral law,
\begin{equation}
F(E) = KE^{-(a+b~\mathrm{Log}E)}
\end{equation}
where $E$ is the energy in keV, $K$ is the photon flux at 1~keV and
the parameter $b$ describes the ``curvature'' of the log-parabola. 
Best fit estimate of the parameters allowed to derive the peak energies
of the spectral energy distributions (SED), $E_p = 10^{(2-a)/2b}$,
that were found at 12 keV and 178 keV for $C_O$ and $C_X$,
respectively, and $b$ was found the same for both components and equal 
to 0.16. 

\subsection{The model at $\gamma$-ray energies}

Observations performed by the COMPTEL and EGRET experiments onboard the 
\emph{Compton Gamma Ray Observatory} (CGRO; Kuiper et~al. 2001; Thompson 
2004) provided light curves above 10 MeV of a good statistical quality,
which show that the pulse shape returns to be similar to that of $C_O$, 
i.e. the optical/soft X-ray one, although some minor differences are 
present, for instance in the shape of P2.
At energies higher than $\sim$500 MeV there is some indication that the 
emission from Ip and P2 appears to increase, in analogy with the X-ray 
band. 
Left panel of Figure \ref{figure1} shows the energy evolution of the P2/P1 flux ratio, 
computed in the phase intervals defined by Kuiper et~al. (2001) that are 
reported in the caption.

To model the $\gamma$-ray emission and the change of the pulse shape, 
MCCM assumed that there are two more, high-energy spectral components, 
$C_{O\gamma}$ and $C_{X\gamma}$, with the same pulse shape of the 
corresponding lower-energy components and with spectral distributions 
also given by Eq. (1). 
This extended model has six new parameters, i.e. the peak energies, 
curvatures and normalizations of the $C_{O\gamma}$ and $C_{X\gamma}$ 
components, that should be determined by data fitting. 
Given the statistical quality of CGRO data, the resulting estimates 
are much more uncertain than in the X-rays. 
Therefore, we assumed that the curvature parameters were equal 
to the $C_O$ and $C_X$ ones ($b = 0.16$), and adjusted the normalizations 
and peak energies to reproduce the observed total (phase-averaged) 
spectrum. 
Peak energies of $C_{O\gamma}$ and $C_{X\gamma}$ were found to lie 
around 300 MeV and 2 GeV, respectively.
To be consistent with the upper limits to the TeV pulsed emission 
(e.g. Lessard et~al. 2000) we added also an exponential cutoff to both 
$C_{O\gamma}$ and $C_{X\gamma}$, at the energy $E_c=15$ GeV. 
This value was not compelled by observational or physical reasons, 
but was simply a guess to take into account the upper limits.
The MCCM model was then able to reproduce both the broadband energy 
spectrum of the total pulse (Figure~\ref{figure1}, right panel), and the spectra of 
the P1, Ip and P2 phase intervals.
In the left panel of Figure~\ref{figure1} we plotted also the fitted and extrapolated 
P2/P1 ratio for two values of the $C_{O\gamma}$ cut-off energy, i.e. 
11 and 15 GeV.
This ratio depends on the normalizations and shapes of the $C_O$ and 
$C_X$ components in the proper phase ranges, and this is not immediately 
apparent in the phase averaged spectrum of Figure \ref{figure1} (see MCCM 
for details).

\section{The pulse shape and spectrum in the GeV band}

\begin{figure}
\resizebox{\hsize}{!}{\includegraphics{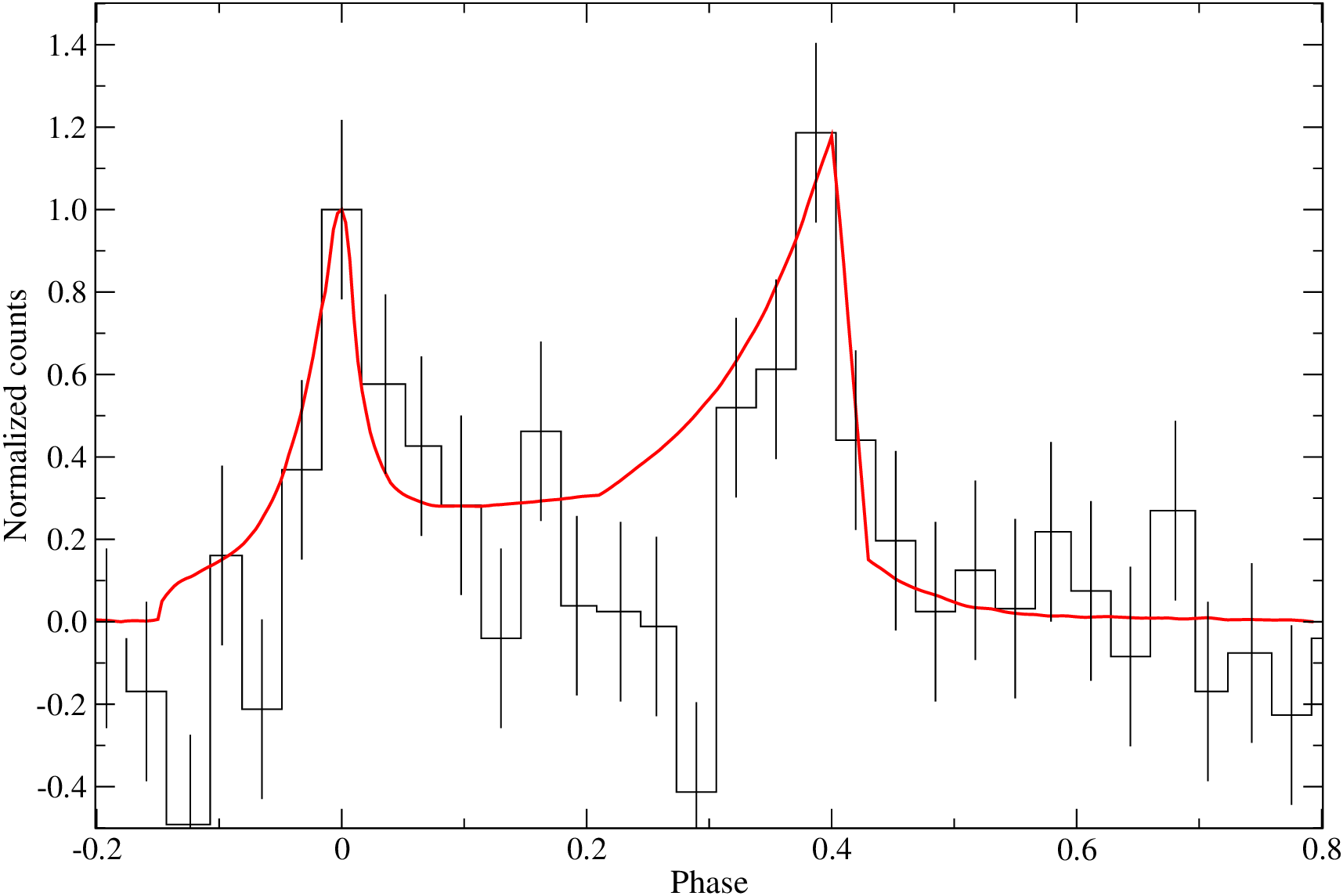}}
\caption{MAGIC observed pulse shape above 25 GeV, normalized at the count 
rate of the first peak, compared with the prediction of MCCM (solid line).}
\label{magic}
\end{figure}

The pulse shape observed above 25 GeV by the MAGIC telescope
(Aliu et al. 2008) is shown in Figure~\ref{magic}, 
where it is normalized to the counts of the first peak. 
Despite the statistical quality of data, is well apparent that 
P2 has an amplitude comparable and possibly higher than P1, 
in agreement with the trend barely apparent from the EGRET data 
(Thompson 2004).

In Figure~\ref{magic} it is also plotted the pulse shape obtained 
from our model, with the same parameters' values used in MCCM, for 
an energy of 25 GeV.
The cut-off energy (15 GeV) is the same for both the $C_{O\gamma}$ 
and $C_{X\gamma}$ components. 
The amplitude of the peaks is very well reconstructed; the pulse 
shape in the MAGIC band predicted by our model has small variations 
with energy, with the height of P2 that varies by about 15\% between 
25 and 60 GeV.

The spectral distribution in the MAGIC range is still poorly determined,
mainly because of the uncertainties in its energy scale.
Aliu et al. (2008) combined the very high energy MAGIC data with the 
COMPTEL and EGRET points and evaluated that a possibile exponential cut-off 
can be at $E_{c} = 17.7 \pm 2.8 \pm 5.0$ GeV, thus compatible to the value assumed 
in the MCCM model. 
However, the CGRO-MAGIC data are compatible with both an exponential or a super-exponential shape, 
with a preference for the former.
We reported the flux estimate in our SED model (Figure~\ref{figure1}, right panel) 
and found that it is in a substantial agreement with the extrapolation which was
introduced to take properly into account the EGRET data.     
Finally, note that in the MCCM model the flux above $\sim$10 GeV 
is mainly due to the $C_{X\gamma}$ component (Figure~2). 
This is not in contrast with the evidence of P1 in the pulse profile, 
because its flux is concentrated in a quite narrow phase interval, 
and $C_{X\gamma}$ has however a wide pedestal extending below P1 
(Figure~\ref{magic}).

The good matching of the predicted pulse shape with the MAGIC one 
implies that the cut-off energy should be similar for $C_{O\gamma}$ 
and $C_{X\gamma}$, because a lower cut-off for $C_{O\gamma}$ would 
imply an higher P2/P1 ratio (as shown in Fig.~\ref{figure1}) and 
therefore a higher amplitude for the second peak.

One of the assumptions of the MCCM model was that the phase 
distributions of the various components do not change with the energy.
It was essentially motivated by the need to have a low number of 
parameters and supported by the satisfactory agreement of the 
computed pulse profiles with data.
However, in Massaro et al. (2000) it was already shown that the 
introduction of a mild energy dependence of two parameters of the analytical description of $C_X$ 
would improve the pulse profile modelling, particularly in the Ip region.     
Our extrapolation at 25 GeV predicts a higher Ip than that given by the MAGIC data, especially 
in the leading wing of P2.
We expect a number of detected events in the Ip phase interval, 
in excess to the off-pulse level, around 2600, 
whereas the measured excess is $\sim$900 counts, but still compatible with the off-pulse emission, due to the large uncertainty.
Note that the S/N ratio does not allow to reach any firm conclusion: there is, for example, a dip at phase $0.3$ (another one is at phase $-0.15$), well 
below the mean off-pulse level, that reduces the Ip content.
The assumption that the shape of $C_{X\gamma}$ is equal to that of
$C_{X}$ could not be exactly verified at such different energies.
They could be different, with a shallower profile for the high-energy 
counterpart, as already suggested in MCCM upon considering the Ip spectrum.

\section{Discussion}

On the basis of a large collection of data, from the optical band
to $\gamma$-ray energies, we developed a model (MCCM) that describes
with a good accuracy the spectral and phase distributions.
This model has essentially an heuristic value, being aimed to
establish a consistent scenario for the development of a realistic 
physical model.
The recent detection of the pulsed emission from Crab at energies
above 25 GeV by the MAGIC collaboration is a relevant result that 
completes our knowledge on the spectral properties of this important 
source.
We have shown here that these results are in a good agreement with 
the expectations from the MCCM model and confirm its validity. 

It would be interesting to investigate theoretically the physical 
plausibility for the presence of two couples of emission components 
producing the observed spectra and phase distributions.
A possibility to be further developed arise from the similar curvature 
parameter for the X-ray and $\gamma$-ray components, that suggests a 
synchrotron-self-Compton (SSC) mechanism for the emission (Morini 1983, 
Cheng \& Wei 1995). 
A stream of relativistic electrons moving along the magnetic field 
lines would undergo a Compton upscattering on the synchrotron photon 
field.
A different location and azimuthal distribution in the magnetosphere 
would explain the different pulse shape for the $C_{O}$ and $C_{X}$ 
components.
An alternative hypothesis is that the $\gamma$-ray components are 
photons emitted by primary electrons via the curvature radiation 
mechanism, and the softer ones are the synchrotron emission from 
secondary pairs.
However, in this case one would expect a more pronounced curvature 
of the log-parabolic spectra for the latter components.
In fact, if the electron spectrum is a log-parabola with a curvature 
parameter $r$, in the $\delta$-approximation the synchrotron curvature 
$b$ would be equal to $r/4$ (Massaro et~al., 2006b), while for 
curvature radiation this parameter would be $r/9$, because of the 
dependence on the electrons' Lorentz factor as $\gamma^{3}$ instead 
of $\gamma^{2}$ for the synchrotron process.
About the log-parabolic (or log-normal) energy distribution of electrons,
we recall that it can originate by stochastic acceleration processes or
when the acceleration probability is energy-dependent.
 
Several models have been proposed to explain the high-energy emission 
from pulsars, both in the outer gap (Cheng et al., 1986, 2000) 
and in the polar/slot gap (Muslimov \& Harding 2003, 2004) frameworks.
A common feature of these models is that the broadband emission 
comes from components originating from different physical processes.

Takata \& Chang (2007) developed a 3D outer gap model based on the 2D 
analytical solution of the accelerating field and particle motion 
by Takata et al. (2004, 2006) and Hirotani (2006).
In this model the X-ray emission of P1 is due to two separate 
components, with curved, roughly log-parabolic, spectra and originating by synchrotron emission 
of secondary pairs in different regions of the outer gap, i.e. below 
and beyond the null-charge surface.
Their explanation of the $\gamma$-ray spectrum invokes other 
two components due to Inverse Compton scattering of secondary pairs and 
curvature radiation from primary electrons.
Their modelling of P2 and Ip, unlike the MCCM model, has different weights 
for these two components.
The increase of P2 and Ip with respect to P1 in this model is due 
to the fact that the emission comes from regions in the lower 
magnetosphere, where the high magnetic field produces harder spectra.

More recently, Harding et al. (2008) performed a 3D simulation of the 
emission from the Crab pulsar, assuming a Slot Gap accelerator,
in which the emission is due to two distribution of particles, 
primary electrons and electron-positron pairs.
The X-ray emission is due to the synchrotron radiation 
from the pairs, while the $\gamma$-rays are curvature 
radiation and the synchrotron radiation resulting from the resonant 
cyclotron absorption of radio beam photons by the primary particles. 

Future observations, in particular with the Fermi Gamma-ray Space Telescope, 
will be very useful to confirm the spectral cut-off observed
by MAGIC, to improve the estimate of $E_c$, and to bridge the gap between the 100--500
MeV EGRET lightcurves and the $\ge$25 GeV MAGIC one.
It will be possible to follow in detail the evolution of the P2/P1 flux ratio and to
verify with a much higher statistics whether the Ip emission is actually related to that of P2.

Another future interesting test will come from phase-resolved X-ray 
polarimetry, where MCCM expect that the hard X-rays polarization properties 
of P2 should become increasingly similar to those of Ip, 
because of the higher contribution of the $C_{X}$ component.

\end{document}